\documentclass[twocolumn]{aastex63}

\usepackage{comment}
\usepackage{cancel}
\usepackage{soul}
\received{October 5, 2021}
\revised{\today}
\accepted{\today}
\submitjournal{ApJL}

\shorttitle{CI Tau SED}
\shortauthors{Muley and Dong}
\begin{document}

\title{CI Tau: A controlled experiment in disk-planet interaction}

\correspondingauthor{Dhruv Muley}
\email{dmuley@berkeley.edu, rbdong@uvic.ca}

\author{Dhruv Muley}
\affiliation{Department of Physics and Astronomy, University of Victoria, 3800 Finnerty Rd., Victoria, BC V8P 5C2, Canada}
\affiliation{Max-Planck-Institut für Astronomie, Königstuhl 17, Heidelberg 69117, Germany}

\author{Ruobing Dong}
\affiliation{Department of Physics and Astronomy, University of Victoria, 3800 Finnerty Rd., Victoria, BC V8P 5C2, Canada}

\nocollaboration{2}



\begin{abstract}
CI Tau is a young (${\sim}2$ Myr) T Tauri system with a substantial near-infrared excess in its SED, indicating that the protoplanetary disk extends very close to its star. This is seemingly at odds with the radial-velocity discovery of CI Tau b, a ${\sim}12 M_J$ planet at $\sim0.1$ au, which would be expected to carve a wide, deep cavity in the innermost disk. To investigate this apparent contradiction, we run 2D hydrodynamics simulations to study the effect of the planet on the disk, then post-process the results with radiative transfer to obtain an SED. We find that at ${\sim}0.1$ au, even such a massive companion has little impact on the near-infrared excess, a result that holds regardless of planetary eccentricity and dust size distribution. Conversely, the observed full-disk signature in CI Tau's SED is consistent with the existence of the hot super-Jupiter CI Tau b. As our simulations uncover, clear transition-disk signatures in SEDs are more likely to be signposts of nascent ``warm'' Jupiters, located at around 1 AU in the future habitable zones of their host stars.

\end{abstract}



\section{Introduction} \label{sec:intro}

The spectral energy distribution (SED) of a planet-forming system consists of three basic components: the stellar photosphere, accretion luminosity, and an excess at infrared and longer wavelengths emitted by dust in the protoplanetary disk. In some cases, the near-infrared (NIR) portion of the disk excess is reduced, implying that dust has been depleted in the innermost, hottest part of the disk. Such SEDs are thought to reflect a ``transitional'' phase in disk evolution \cite[e.g.,][]{Calvet2002,Calvet2005} where they clear from the inside out, a picture increasingly borne out in recent years by high-resolution observations \citep[e.g.,][]{Andrews2011,Espaillat2014,VanDerMarel2018}

Given their ubiquity, planets are expected to play a key role in carving transition cavities, and conversely, transition cavities are thought to be signposts of planet formation. These prospects have received substantial theoretical \citep{DodsonRobinson2011,Zhu2011,Zhu2012,dong12cavity,Muley2019,Bae2019} and observational \citep{Sallum2015,Keppler2018} interest. However, planets are difficult to directly image in transition disks, and often have large, model-dependent uncertainties in mass that make their role in carving a given cavity unclear \citep[see ][and references therein]{Muller2018,AsensioTorres2021}. 

In this context, the ${\sim}2$ Myr old CI Tauri system is of particular interest. It is one of the rare protoplanetary-disk systems where a super-Jupiter planet candidate has been detected independently of imaging---in this case, by radial velocity \citep{JohnsKrull2016}. The detection of CI Tau b is not uncontroversial at the moment \citep{Donati2020}. The companion is predicted to have a mass of ${\sim}12 M_J$, a semimajor axis of $0.08$ au, and an eccentricity of $0.25 \pm 0.15$ \citep{Flagg2019,Donati2020}. However, the NIR excess of the system is characteristic of a full, unperturbed disk, with no obvious signs of disk-planet interactions \citep[e.g.,][]{McClure2013,McClure2013b}. Determining that the cavity excavated by CI Tau b would not substantially impact the SED would resolve this apparent contradiction. This would provide a starting point to test which disk and planet parameters \textit{do} impact the SED, and more broadly, under what circumstances transitional disk-like SEDs can be seen as signposts of inner massive planets. The existence of hot Jupiters in $\sim$Myr old systems would pose a direct test of formation models such as \textit{in-situ} growth or disk migration \citep{Dawson2018}.

To this end, we carry out hydrodynamical simulations of disk-planet interactions, then post-process the results with a radiative-transfer code, in order to obtain disk structures and SEDs. We vary parameters such as planetary semimajor axis, eccentricity, and dust size distribution. By focusing on cavities $\lesssim 5$ au in width extending all the way to the star, and by self-consistently importing a dust distribution from the hydrodynamics, our study complements previous work by \cite{Zhu2012} and \cite{Dong2012}, who used semi-analytic modeling to study the SED of disks with ${\sim}20$ au and ${\sim}65$ au cavities respectively. Section \ref{sec:methods} contains our methods, Section \ref{sec:results} our results, and Section \ref{sec:conclusion} our conclusions.

\section{Methods} \label{sec:methods}
\subsection{Full-disk case} \label{sec:methods_full}

Before investigating the effect of a planet, we first use the \texttt{HOCHUNK3D} Monte Carlo ray-tracing code \citep{Whitney2013} to find an unperturbed, analytical disk model that best reproduces the observed photometric SED. We largely follow the CI Tau model in \citet{Ballering2019} in this step. We use a photospheric temperature of $T_* = 4000$ K and a total luminosity of $L_{\rm tot} = 1.18 L_\odot$ \citep{Andrews2013};\footnote{Corrected for the fact that \cite{Andrews2013} used a distance of 140 pc for CI Tau instead of the Gaia distance of 158 pc.} we assume that 75\% of this luminosity ($L_* = 0.89 L_\odot$) comes from direct photospheric emission, while 25\% ($L_{\rm acc} = 0.30 L_\odot$) comes from accretion onto the star, in order to match the modeling work of \cite{McClure2013} at $\lambda < 1 \mu \rm{m}.$ The photospheric luminosity corresponds to a stellar radius of $R_* = 1.96 R_\odot$, typical of T Tauri stars. Using this radius, and using a stellar mass of $M_* = 0.9 M_\odot$ \citep{Guilloteau2014,Flagg2019} we estimate a mass accretion rate of $\dot{M}_{\rm acc} = 2.09 \times 10^{-8} M_\odot \textup{yr}^{-1}$, within order-unity of that in \cite{McClure2013}. As they did, we model the accretion luminosity as an 8000 K blackbody emanating evenly from the stellar surface, and find likewise that we reproduce the observed SED in the optical. We use $5 \times 10^8$ photon packets to perform ray-tracing.

\begin{figure}
    \centering
    \includegraphics[width=0.5\textwidth]{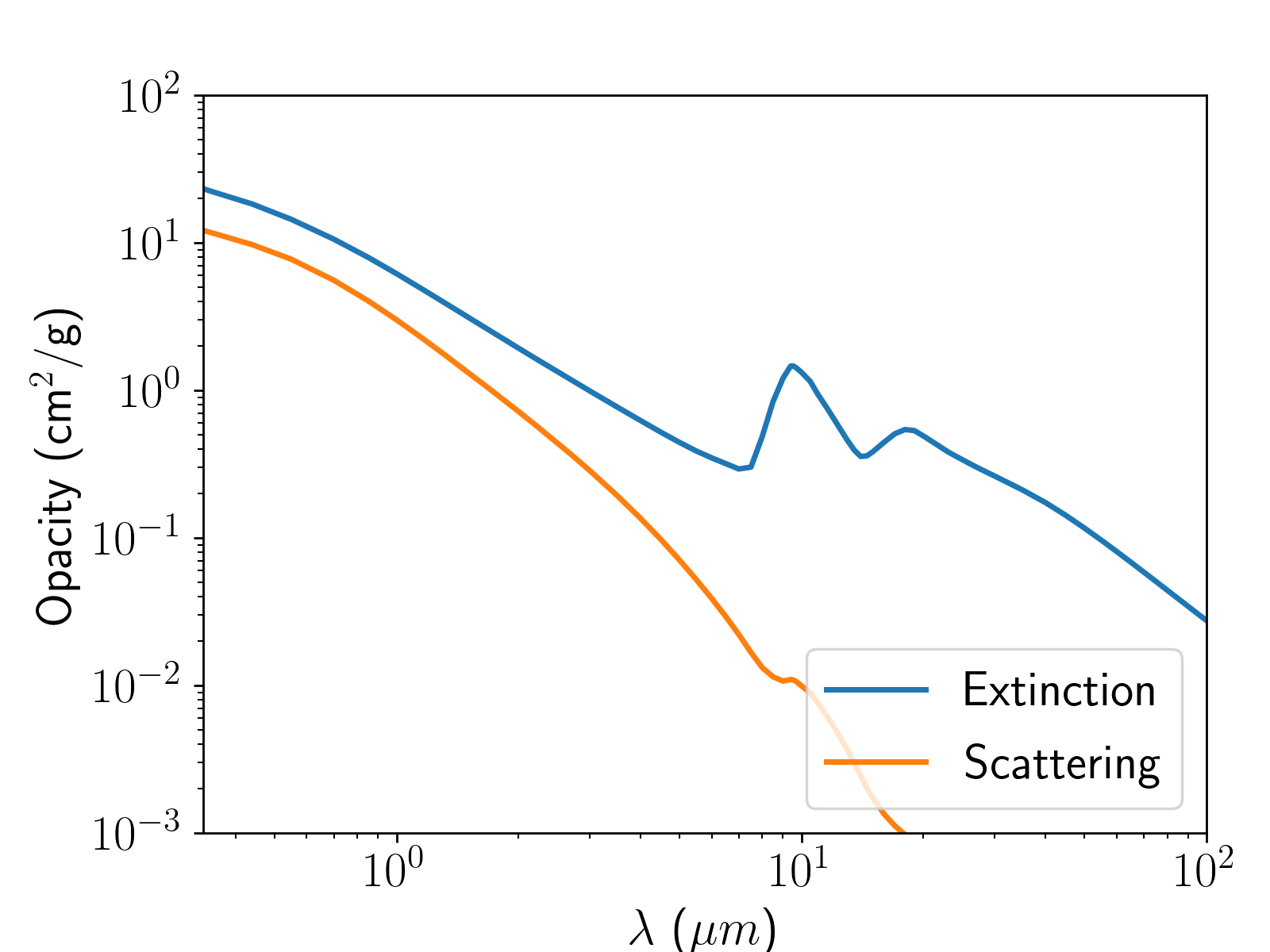}
    \caption{Opacity for our dust prescription, at a typical dust-to-gas ratio of 0.01 with 10\% small grains by mass. The lack of large grains means that the optical thickness and emission of our disk at wavelengths longer than $\sim10\mu$m, which we do not focus on, is understated.}
    \label{fig:opacity}
\end{figure}

Our models use \cite{Kim1994} small interstellar medium grains (hereafter ISM), which provide a good fit to the SED photometry for wavelengths $\lambda \lesssim$100$\mu {\rm m}$, as well as the 10 $\mu m$ feature observed by Spitzer/IRS. Such grains, with Stokes numbers ${\rm St} \ll 1$, are well-coupled to the gas, and so do not require separate modeling in our hydrodynamics simulations (Section \ref{sec:hydro}). We use a dust sublimation temperature $T_{\rm sub} = 1700$ K, with a sublimation radius at $r_{\rm sub} = 0.08$ au and a curved wall with $r_{\rm curve} = 0.003$ au to account for pressure-mediated vertical dependence in the sublimation radius \citep{Isella2005}. We assume a dust-to-gas ratio of 0.01, and fiducially assign 10\% of this mass to small grains \citep{Dong2012}; we plot the relevant opacity in Figure \ref{fig:opacity}. The rest of the dust mass is assumed to be in bigger grains. Because our focus is on the NIR excess, we need not model this population of larger grains, which supply only a negligibly small fraction of total opacity at $\lambda \lesssim 10 \micron$.

Our disk is flared, with an aspect ratio of 
\begin{equation}\label{eq:sch}
    H(r) \equiv c_s(r)/v_k(r) = H_0 (r/1 {\rm \ au})^{\beta}
\end{equation}
where $c_s$ is the sound speed, $v_k$ the Keplerian velocity, $r$ the distance from the star, $H_0 = 0.065$ the aspect ratio at 1 au, and $\beta = 0.125$ the flaring index. 
These parameters are constrained by fitting the disk SED \citep{Ballering2019}. The disk profile thus does not reflect that obtained by simply taking the radiative transfer output temperature and assuming hydrostatic equilibrium. Such parametrizations are often adopted in disk modeling \citep[e.g.,][]{Andrews2011,Dong2012,Zhang2014}, as certain physical processes that affect the thermal state of the disk, including planetary shocks and viscous heating in a deep cavity, are not readily and self-consistently treated in radiative transfer codes.

We use a total (gas and dust) mass of $M_{disk} = 0.02 M_\odot$ inside its outer boundary $r_{\rm out} = 100$ au. The surface density scales as 
\begin{equation}\label{eq:dens}
    \Sigma(r) = \Sigma_0 (r/1 {\rm \ au})^{-1} \exp(-(r/r_{\rm trunc})^{-2})
\end{equation}
where $\Sigma_0 = 283 {\rm \ g}/{\rm cm}^2$ and $r_{\rm trunc} = 0.08$ au---where the rotation period of the star (and therefore, that of its frozen-in magnetic field) coincides with the Keplerian period, leading to a magnetospheric truncation of the disk. This happens to coincide with the dust sublimation radius.

\begin{figure}
    \centering
    \includegraphics[width=0.425\textwidth]{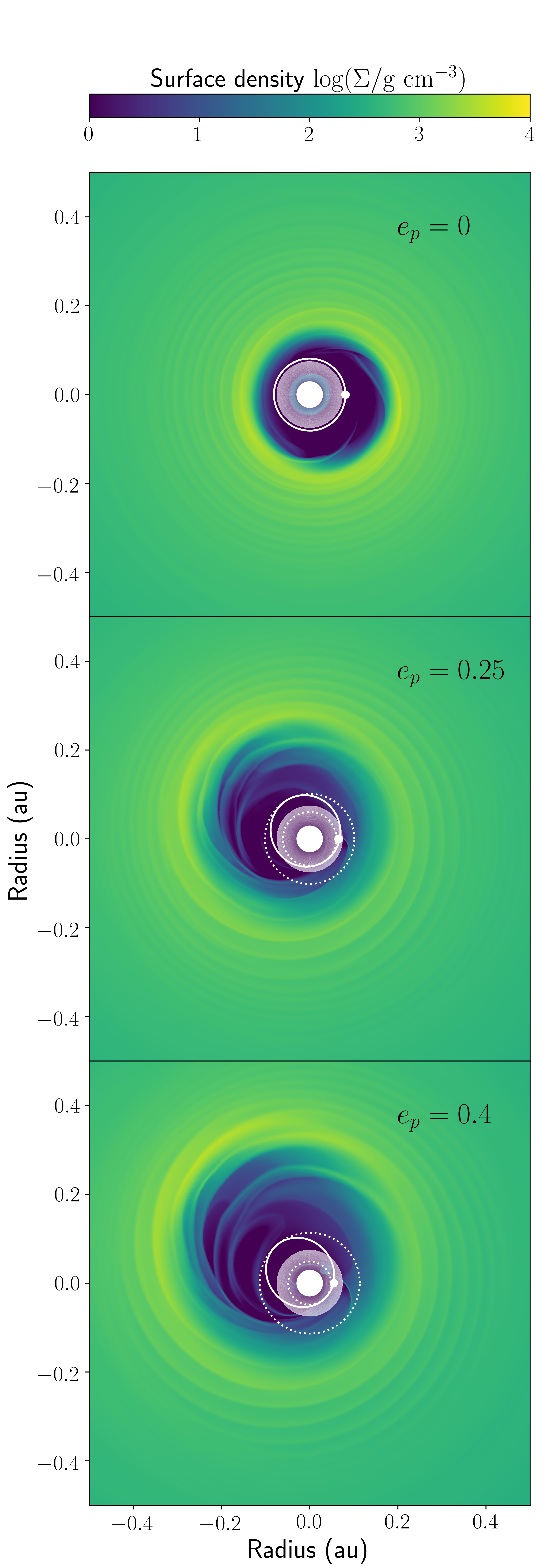}
    \caption{Surface density in all simulations at $t = $ 1143 orbits. White dots indicate planet location; solid lines its orbit; and dotted lines the region between perihelion and aphelion, which is subject to accretional clearing. The faded region is inside the sublimation radius $r_{\rm sub} = 0.08$ au; we include it for accuracy in our hydrodynamics, but in itself it makes no contribution to the SED.
    }
    \label{fig:2d_sigma}
\end{figure}

\subsection{Planet-carved cavity case}\label{sec:hydro}
To simulate the effect of CI Tau b on its disk, we run 2D hydrodynamics simulations with \texttt{PEnGUIn} \citep{Fung2015}, a GPU-accelerated, Lagrangian-remap code that employs the piecewise parabolic method (PPM) of fluid reconstruction. As in \cite{Muley2021}, we solve the viscous, compressible Navier-Stokes equations, with gravitational forces from both the star and the planet. We relax the local temperature in each grid cell to the background temperature on a timescale $t_c = 10^{-4}$ local dynamical times, making our simulations effectively locally isothermal \citep{MirandaRafikov2020}. This facilitates comparison with previous simulations of super-Jupiters \citep[e.g,][]{Dunhill2013,Ragusa2018,Muley2019,Teyssandier2020}, and simplifies the complex thermal physics \citep{Rafikov2016} and gas-grain coupling issues \citep{Bae2021} that would emerge for the strong shocks they would generate.

Per \cite{Flagg2019}, we set CI Tau b's mass $M_p = 11.6 M_J$ and the stellar mass $M_* = 0.9 M_\odot$. We use a semimajor axis $a_p = 0.08$ au, and test eccentricities of $e_p = \{0, 0.25, 0.4\}$, bracketing the $e_{\rm p, obs} = 0.25 \pm 0.15$ range expected from observations. These parameters are fixed throughout the simulation, and do not change due to the gravitational back-reaction of disk material onto the planet; however, similar values are naturally achievable for super-Jupiter planets when the planet is free to migrate \citep{Papaloizou2001,Bitsch2013,Dunhill2013,Ragusa2018,Muley2019}.

We allow the planet to accrete disk material according to the prescription of \cite{Muley2019}, at a rate 
\begin{equation}
    \dot{\Sigma}(r, \phi)_{\rm acc} = -\frac{\Sigma(r, \phi)}{k t_{\rm ff}}
\end{equation}
where $k = 10$, the typical free-fall timescale $t_{\rm ff} = \sqrt{2r_{\rm acc}^3/GM_p}$, and the typical accretion radius $r_{\rm acc} \equiv r_{\rm H}/3 = (r_p/3) (M_p/M_*)^{1/3}$ ($r, \phi$ are cylindrical coordinates of a grid cell within the disk, while $r_p$ is the instantaneous radial location and $r_{\rm H}$ is the Hill radius of the planet). This rate is limited, spatially, by a Gaussian function centered on the planet:
\begin{equation}
    \dot{\Sigma}(r, \phi)_{\rm max} = \frac{\dot{M}_{\rm max}}{2\pi r_{\rm acc}^2} \exp{\left(-\frac{r^2 + r_p^2 - 2rr_p \cos\phi'}{2r_{\rm acc}^2}\right)}
\end{equation}
where we pick $\dot{M}_{\rm max} = 25 M_J{\rm /yr}$, allowing essentially unlimited accretion within the planet's Hill radius. The accreted mass is removed from the domain entirely and is not added to that of the planet.

We use a Shakura-Sunyaev viscosity prescription $\nu = \alpha c_s h$ with $\alpha = 10^{-3}$, a value commonly used in hydrodynamical simulations of protoplanetary disks
%to suppress the growth of Rossby vortices 
\citep[e.g.,][]{Fung2014,Fung2016,Bae2019} and emerging naturally from small-scale turbulence in nearly-isothermal disks such as the one we use \citep{Manger2021}. Following \citet{Ballering2019} we set the background temperature equal to that implied by Equation \ref{eq:sch}, scaling as $T \propto r^{-0.75}$ (note that the disk is being heated by both stellar irradiation and accretion). The initial surface density is given by Equation \ref{eq:dens}, augmented with an additional exponential truncation factor $\exp(-(r/r_{\rm out})^2)$ for $r_{\rm out} = 2.5$ au, to minimize spurious addition and removal of mass and angular momentum at the (circularly symmetric) outer boundary. As shown later, we mainly focus on the inner disk structure within 1 AU and photometry at wavelengths shorter than $10\micron$.

We run hydrodynamics simulations to 25 years, or 1143 orbits of CI Tau b, long enough to establish a quasi-steady state. Our domain spans a radial range of $\{0.03, 5\}$ au, and the full $2\pi$ in azimuth; we use a zero-gradient outflow condition at the inner boundary, and a fixed condition at the outer one. Numerical tests (not shown here) reveal that shifting the inner boundary inward by a factor of 2 does not impact cavity depth. We cover this range with $1000 \ (r) \times 1228 \ (\phi)$ cells, logarithmically spaced in radius and evenly spaced in azimuth, yielding an effective resolution of ${\sim}$9 cells per scale height at the location of the planet. This resolution is similar to that of the CI Tau simulations in \cite{Teyssandier2020}, which in turn is based on detailed numerical tests by \cite{Teyssandier2017}.

For our radiative-transfer modeling, we rescale the innermost part of our grid such that $a_p = \{1, 10^{0.5}, 10\} \times 0.08$ au, and cut out the innermost cells up to $r_{\rm sub} = 0.08$ au. For computational efficiency, we downsample the hydrodynamics grid by a factor of 2, and pad the remainder of the 100 au domain with surface density matched at the hydrodynamics cut and scaling as $r^{-1}$; we normalize in each case such that disk mass always equals the fiducial $M_{disk} = 0.02 M_\odot$. Subsequently, we vertically ``puff up'' the 2D disk assuming local hydrostatic equilibrium as a Gaussian, with standard deviation given by Figure \ref{eq:sch}. This gives us a 3D polar grid with resolution $\{334, 358, 348\} (r) \times 614 (\phi) \times 108 (\theta)$ for $a_p = \{1, 10^{0.5}, 10\} \times 0.08$ au.
In the vertical direction, we cover a range of $\theta = \{\pi/2 - 0.5, \pi/2 + 0.5\}$, spacing cells parabolically so $\Delta \theta(\theta) = |\theta - \pi/2| (0.5^2)/108$.

\begin{figure}
    \centering
    \includegraphics[width=0.5\textwidth]{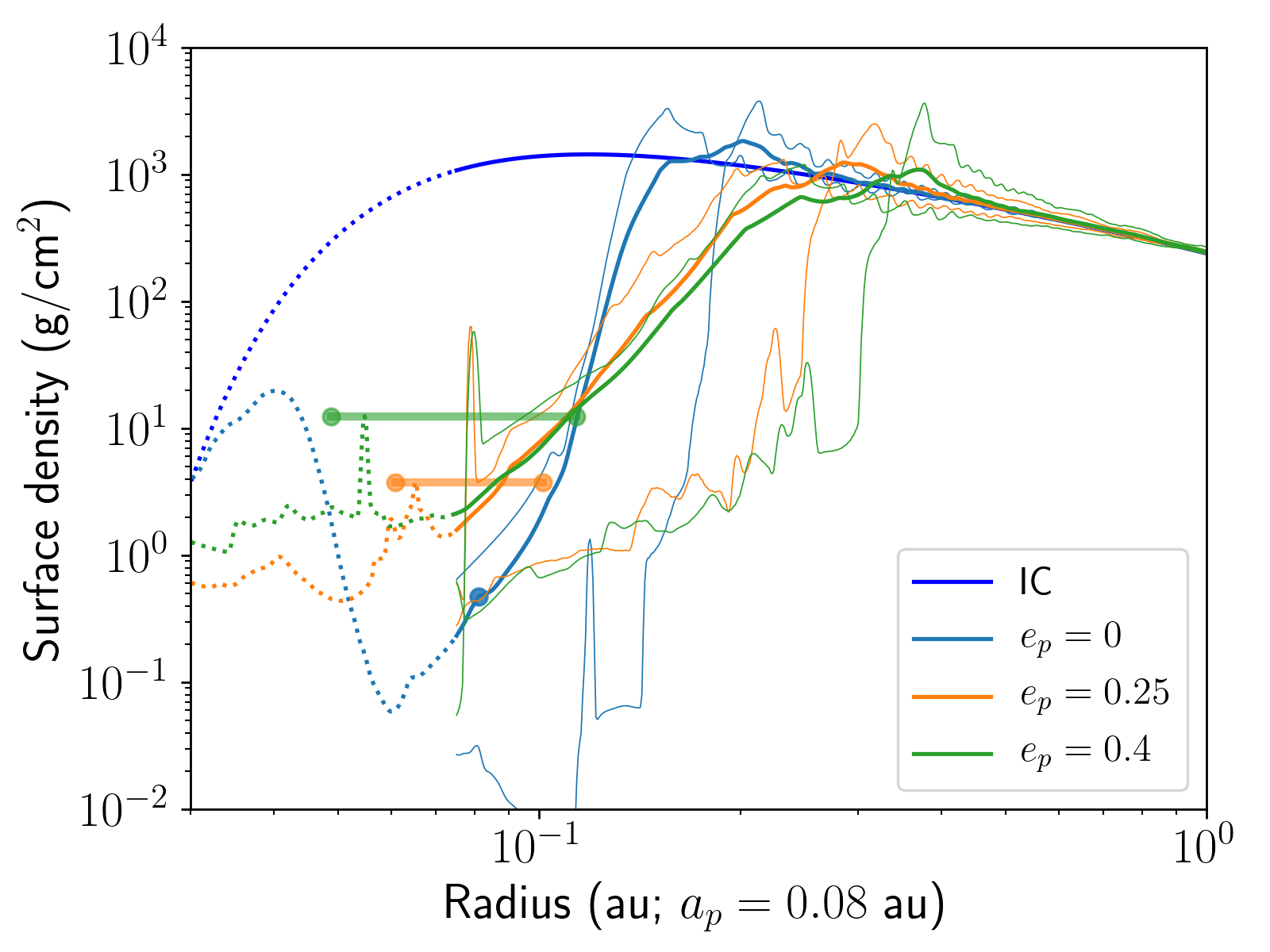}
    \caption{Surface density in the innermost 1 au of our hydrodynamical simulations, assuming an $a_p = 0.08$. Thick lines are azimuthally averaged, while thin lines are radial cuts of the closest and farthest extent of the cavity wall. Dotted lines are inside the dust sublimation radius, and therefore make no contribution to the SED. Transluscent bars indicate the perhielion-aphelion range of the planet for each eccentricity. As expected, an eccentric planet creates a wider, and more eccentric, gap.
    }
    \label{fig:sigma_comparison}
\end{figure}

\section{Results} \label{sec:results}

\subsection{Hydrodynamics}
Planets open gaps in disks by the action of ordinary Lindblad torques on the surrounding disk material. When the planet-star mass ratio $q \equiv M_p/M_* \gtrsim 0.003$, a secondary effect---the eccentric Lindblad torques---become prominent and excite eccentricity in the disk \citep{Kley2006}. Gas parcels with large semimajor axes can thus intersect the planet's orbit and get consumed by accretion, leaving a wider gap than ordinary Lindblad torques alone would indicate. These effects are enhanced substantially when the planet is itself eccentric and exposed to a wider swathe of the disk, as visible in Figure \ref{fig:2d_sigma}. This is similar to the result in \cite{Muley2019}, but the locking of disk and planet pericenters means the disk remains eccentric in our present work, rather than circularizing as in theirs. Our simulations also show ``horseshoes'', azimuthal asymmetries in surface density formed by companions with mass ratios $q \gtrsim 0.01$ \citep{Ragusa2020}.

In Figure \ref{fig:sigma_comparison}, we show more quantitative disk surface density profiles for all the eccentricities we test, with radial cuts of the closest and farthest approaches of the cavity edge. Our plot assumes the fiducial $a_p = 0.08$ au, corresponding to CI Tau's orbital parameters; we note that maintaining a disk mass of $M_{\rm disk} = 0.02 M_\odot$ means that plotted densities scale down, and distances scale up, by factors of $10^{0.5}$ ($a_p = 0.26$) and 10 ($a_p = 0.81$). For large eccentricities, the cavity width is up to 5$\times$ the planetary semimajor axis at its widest point, and can be depleted by a factor of ${\sim}10^2-10^3$; in the zero-eccentricity case, the cavity is only about $2.5\times$ as wide as the planetary, but deeper than in the eccentric case. Our cavity depths are broadly consistent with those of \cite{Teyssandier2020}, who focused on pulsed accretion, also observed here. However, we concentrate on gap depths, so we use a zero-gradient inner boundary that allows material to flow into and out of the domain. It is not as well-suited to studying pulsed accretion as their ``diode'' boundary, which only allows material to exit; this enforces an $\dot{M}_{\rm disk} \leq 0$ at all times, at the cost of some excess cavity depletion.

\begin{figure}
    \centering
    \includegraphics[width=0.5\textwidth]{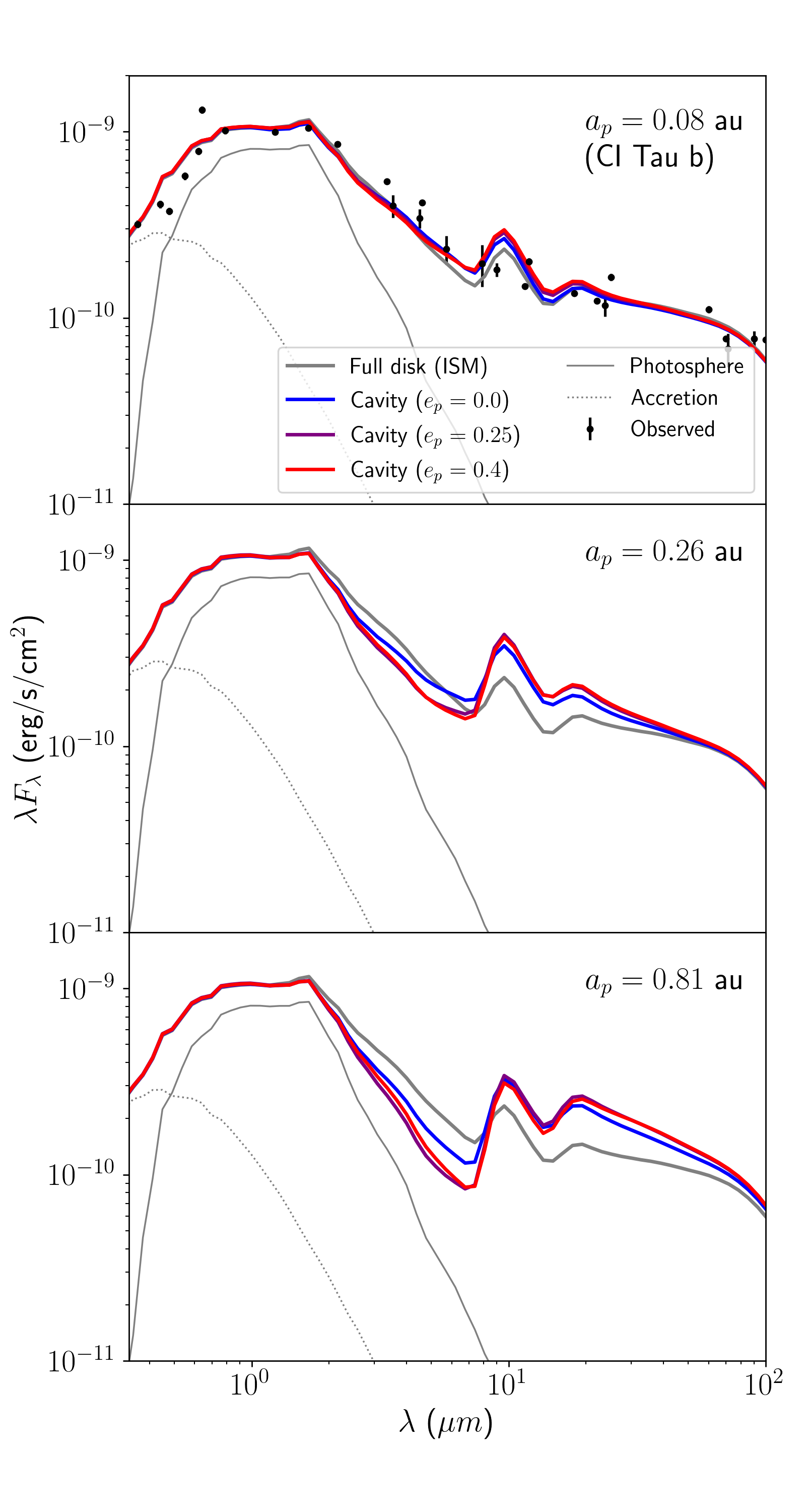}
    \caption{A comparison between SED photometry, our full-disk model (the same in all panels), and planet-carved gap for various eccentricities and planetary semi-major axis $a_p$. In the upper panel $a_p=0.08$ AU, as in the case of CI Tau. In the other two panels we explore cavities opened by planets with larger $a_p$. The planet has little effect for $a_p = 0.08$ au, but clearly reduces NIR excess for $a_p = 0.81$ au. In the latter case, the reduction is more pronounced for $e_p = 0.25, 0.4$ than for $e_p = 0$.
    }
    \label{fig:ecc_comparison}
\end{figure}

\subsection{Radiative transfer}
In Figure \ref{fig:ecc_comparison}, we plot model SEDs for the three planetary eccentricities we test, along with a full-disk SED, and compare them with observed photometry points \citep{Andrews2013}. For the fiducial cavity ($a_p = 0.08$ au, leading to $r_{\rm cav} \approx 0.2-0.4$ au, increasing with eccentricity), shown in the upper panel, eccentricity has limited observational impact, with the full disk and $e_p = \{0, 0.25, 0.4\}$ cases all agreeing with each other, as well as the observed CI Tau SED at $\lambda\lesssim10\micron$. We conclude that the existence of a companion in CI Tau is compatible with the system's SED.

\cite{Zhu2012} used hydrodynamics and radiative transfer to study transition cavities between ${\sim}2-20$ au. We connect to their parameter space by rescaling our simulation domain such that $a_p = 0.26, 0.81$ au. The order-of-magnitude range comfortably accommodates not only changes in planetary position, but also any gap-width variations that might be expected due to variations in scale height or viscosity \citep{Dong2017}. When we do so, the differences between the full-disk and planetary cases become more pronounced than for $a_p = 0.08$. Given the opacities in Figure \ref{fig:opacity}, and the surface densities in Figure \ref{fig:sigma_comparison} (as rescaled for the planet location), planetary depletion is sufficient to make the deepest parts of the cavity optically thin; furthermore, it lowers the altitude of the radial $\tau = 1$ surface, meaning cavity material is less irradiated \citep[e.g.,][]{Dong2015}. Both of these effects contribute to a weaker NIR excess at $\lambda \lesssim 10\micron$, while the latter opens up the cavity wall to direct stellar irradiation, creating an excess at $10\micron \lesssim \lambda \lesssim 100\micron$; the precise cutoffs shift outward with increasing cavity radius. Effects are somewhat stronger for $e_p = 0.25, 0.4$, on account of the wider cavity in those cases, than for $e_p = 0$. We stress that all of these results were obtained with an accreting, massive ${\sim}12 M_J$ companion, and thus represent a best-case scenario for carving cavities with transitional disk-like SEDs.
\begin{figure}
    \centering
    \includegraphics[width=0.5\textwidth]{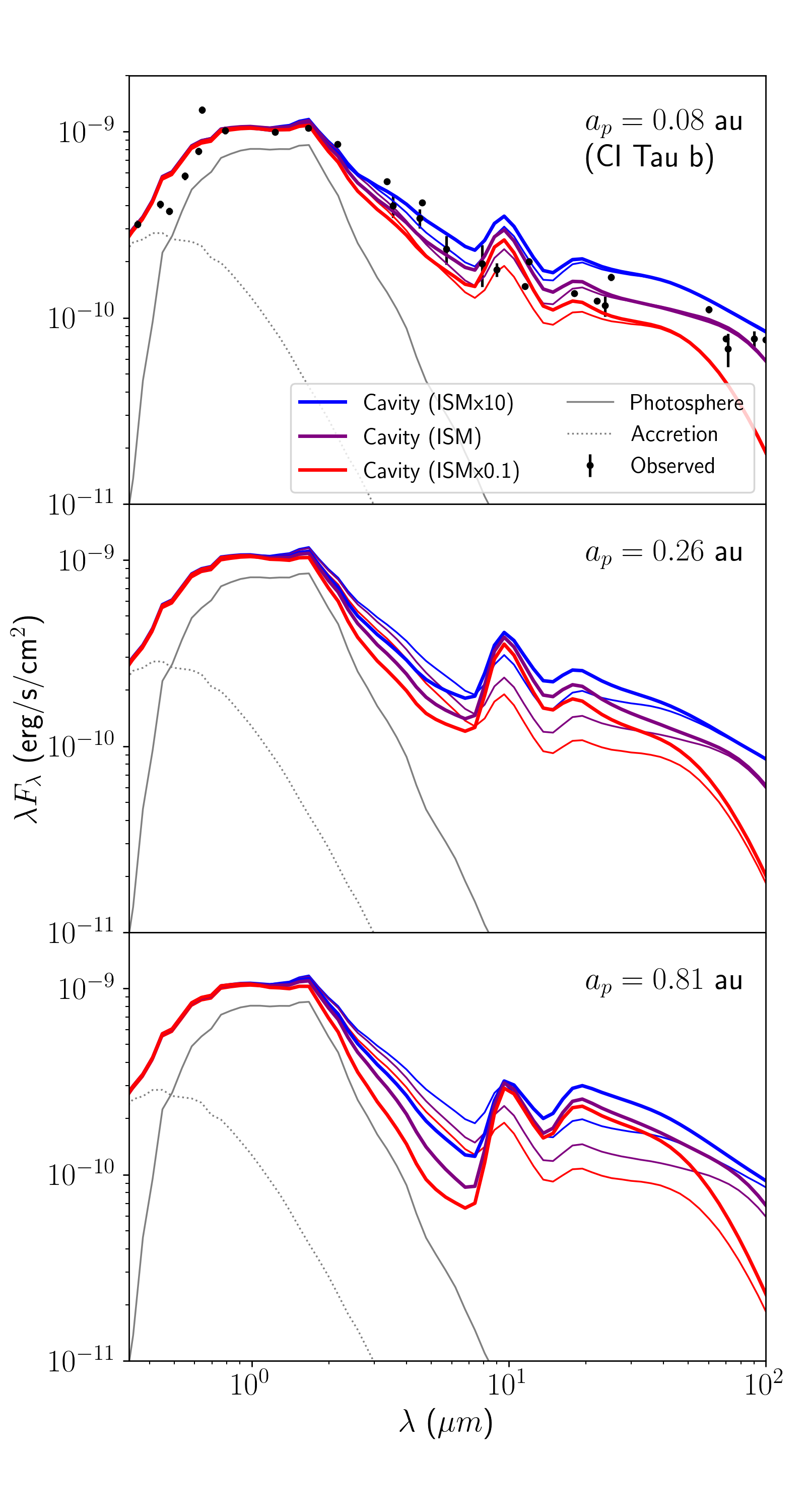}
    \caption{Test of different small-dust-to-gas fraction; different colors correspond to $10\times$, $1\times$, and $0.1\times$ our fiducial small-dust-to-gas ratio of 0.001.
    All planet models have an $e_p = 0.4$. In the upper panel $a_p=0.08$ AU, as in the case of CI Tau. In the other two panels we explore cavities opened by planets with larger $a_p$. 
    In each case, we give a comparison to the expected full-disk model in the same color, but with a thin line. The lower the fraction of small grains, the more pronounced the transition-disk cavity's signature.
    }
    \label{fig:kmh_distance}
\end{figure}

Besides planetary parameters, dust properties also play a role in transition-disk cavity visibility. Coagulation of dust into larger-sized grains reduces the effective surface area per unit mass, and thus opacity, while fragmentation does the reverse \citep{Birnstiel2012}. In our radiative-transfer models, we parametrize these effects by scaling the small dust abundance to $10\times$, $1\times$ and $0.1\times$ its fiducial ratio; this is equivalent to assigning 100\%, 10\%, and 1\% of dust mass to small grains, respectively. Alternatively, these rescalings provide a computationally inexpensive way to bracket the effects that a change in gap depth---and thus, viscosity and scale height \cite{Fung2014} ---might have on the SED. As elsewhere in this work, we do not include emission from large grains. 

We plot our results in Figure \ref{fig:kmh_distance} for the full disk and all of $a_p = \{0.08, 0.26, 0.81\}$ au, choosing the $e_p = 0.4$ models as they represent the best-case scenario for transition-cavity formation. Given the same fractional cavity depletion, the altitude of the radial $\tau = 1$ surface is more strongly reduced, and more of the cavity becomes optically thin, when small grains are fewer to begin with. As expected, for $a_p = 0.08$, the observational impact of this is minimal, but for $a_p = 0.81$, a small-dust fraction $0.1\times$ our fiducial value can lead to a transition disk-like dip that almost fully exposes the bare stellar photosphere. The same process increases the fraction of the cavity wall that is directly irradiated, creating a larger proportional increase in the excess at $10-100 \micron$.

\section{Conclusion} \label{sec:conclusion}
Inspired by the CI Tau system, we use hydroynamical simulations and radiative-transfer post-processing to study the properties of cavities carved by closely orbiting super-Jupiters in protoplanetary disks. All of our simulations use the same, freely accreting $11.6 M_J$ companion orbiting a 0.9 $M_\odot$ star. We test three different orbital eccentricities, $e_p = \{0, 0.25, 0.4\}$. While the fiducial model mimics the CI Tau system with a planet at 0.08 au, we experiment 
%assume a total disk mass $M_d = 0.02 M_\odot$, 
with different semimajor axis and dust-to-gas ratio by testing $a_p = \{1, 10^{0.5}, 10\} \times 0.08$ au and $f_{\rm SD} = \{0.1, 1, 10\} \times 10^{-3}$, with respect to the fiducial values. Our SED modeling builds on previous work \cite[e.g.,][]{Zhu2012,Dong2012,Muley2019} both by studying cavities $\lesssim4$ au in radius, and by self-consistently taking dust surface density from the hydrodynamics rather than parametrizing it with piecewise analytic functions.

By design, our setup represents an extremely optimistic scenario for transition-cavity formation. Our cavity is depleted by a factor of ${\sim}10^{-4}-10^{-3}$ and excavated to a width of ${\sim} 2.5-5\times$ the planet's orbit, the larger figures contingent on the existence of planetary eccentricity. For a planetary semimajor axis ${\sim}1$ au, this makes the deepest part of the cavity optically thin, while lowering the radial $\tau = 1$ surface to expose the cavity wall. This, in turn, creates the characteristic transition disk-like dip from $1-10 \micron$, and an additional excess at $\gtrsim10 \micron$, which become more pronounced for high planetary eccentricities and low initial fractions of small dust. While the same mechanisms are at play for a closer semimajor axis of ${\sim}0.1$ au, the $100\times$-smaller surface area of the excavated cavity means that the NIR excess is indistinguishable from that of an unperturbed disk---a result that holds even in tests with a cavity artificially depleted by a factor of $10^{-6}$ (not shown here). We thus infer that the SED photometry of the CI Tau system is consistent with the existence of the close-in, multi-Jupiter-mass companion, CI Tau b. 

While transitional and pre-transitional SED features are not formed by hot Jupiters \textit{per se}, they are good signposts for nascent ``warm'' super-Jupiters in the eventual habitable zones of their host stars. Future theoretical studies of such systems may sample more finely in time, in order to better understand observed variability in near-infrared excess \citep[e.g.,][]{Espaillat2011,Flaherty2012}. Additionally, a coupled dust-gas code would accurately model large-grain dynamics in such systems, and allow radiative transfer to reproduce the long-wavelength part of the SED. Observationally, such systems may be targeted for radial-velocity surveys if appropriately inclined, or for interferometric observations that reveal the telltale eccentric cavity \citep[e.g.,][Fig. 2 in this work]{Kraus2013}. Beyond individual systems, comparing the populations of disks with transitional SED features to those of warm Jupiters---as  \cite{vanDerMarel2021b} did for resolved transition disks and massive planets more generally---would help elucidate formation mechanisms for these systems. 

\begin{acknowledgments}
{\bf We thank the referee for a prompt and constructive report}. We thank Nienke van der Marel and Shangjia Zhang for useful discussions, as well as Jeffrey Fung for providing us the \texttt{PEnGUIn} code. Computations were performed on infrastructure provided by WestGrid and Compute Canada. R. D. is supported by the Natural Sciences and Engineering Research Council of Canada and the Alfred P. Sloan Foundation.

\end{acknowledgments}

\typeout{}

\bibliographystyle{aasjournal}

\begin{thebibliography}{}
\expandafter\ifx\csname natexlab\endcsname\relax\def\natexlab#1{#1}\fi
\providecommand{\url}[1]{\href{#1}{#1}}
\providecommand{\dodoi}[1]{doi:~\href{http://doi.org/#1}{\nolinkurl{#1}}}
\providecommand{\doeprint}[1]{\href{http://ascl.net/#1}{\nolinkurl{http://ascl.net/#1}}}
\providecommand{\doarXiv}[1]{\href{https://arxiv.org/abs/#1}{\nolinkurl{https://arxiv.org/abs/#1}}}

\bibitem[{{Andrews} {et~al.}(2013){Andrews}, {Rosenfeld}, {Kraus}, \&
  {Wilner}}]{Andrews2013}
{Andrews}, S.~M., {Rosenfeld}, K.~A., {Kraus}, A.~L., \& {Wilner}, D.~J. 2013,
  \apj, 771, 129, \dodoi{10.1088/0004-637X/771/2/129}

\bibitem[{{Andrews} {et~al.}(2011){Andrews}, {Wilner}, {Espaillat}, {Hughes},
  {Dullemond}, {McClure}, {Qi}, \& {Brown}}]{Andrews2011}
{Andrews}, S.~M., {Wilner}, D.~J., {Espaillat}, C., {et~al.} 2011, \apj, 732,
  42, \dodoi{10.1088/0004-637X/732/1/42}

\bibitem[{{Asensio-Torres} {et~al.}(2021){Asensio-Torres}, {Henning},
  {Cantalloube}, {Pinilla}, {Mesa}, {Garufi}, {Jorquera}, {Gratton}, {Chauvin},
  {Szulagyi}, {van Boekel}, {Dong}, {Marleau}, {Benisty}, {Villenave},
  {Bergez-Casalou}, {Desgrange}, {Janson}, {Keppler}, {Langlois}, {Menard},
  {Rickman}, {Stolker}, {Feldt}, {Fusco}, {Gluck}, {Pavlov}, \&
  {Ramos}}]{AsensioTorres2021}
{Asensio-Torres}, R., {Henning}, T., {Cantalloube}, F., {et~al.} 2021, arXiv
  e-prints, arXiv:2103.05377.
\newblock \doarXiv{2103.05377}

\bibitem[{{Bae} {et~al.}(2021){Bae}, {Teague}, \& {Zhu}}]{Bae2021}
{Bae}, J., {Teague}, R., \& {Zhu}, Z. 2021, \apj, 912, 56,
  \dodoi{10.3847/1538-4357/abe45e}

\bibitem[{{Bae} {et~al.}(2019){Bae}, {Zhu}, {Baruteau}, {Benisty}, {Dullemond},
  {Facchini}, {Isella}, {Keppler}, {P{\'e}rez}, \& {Teague}}]{Bae2019}
{Bae}, J., {Zhu}, Z., {Baruteau}, C., {et~al.} 2019, \apjl, 884, L41,
  \dodoi{10.3847/2041-8213/ab46b0}

\bibitem[{{Ballering} \& {Eisner}(2019)}]{Ballering2019}
{Ballering}, N.~P., \& {Eisner}, J.~A. 2019, \aj, 157, 144,
  \dodoi{10.3847/1538-3881/ab0a56}

\bibitem[{{Birnstiel} {et~al.}(2012){Birnstiel}, {Andrews}, \&
  {Ercolano}}]{Birnstiel2012}
{Birnstiel}, T., {Andrews}, S.~M., \& {Ercolano}, B. 2012, \aap, 544, A79,
  \dodoi{10.1051/0004-6361/201219262}

\bibitem[{{Bitsch} {et~al.}(2013){Bitsch}, {Crida}, {Libert}, \&
  {Lega}}]{Bitsch2013}
{Bitsch}, B., {Crida}, A., {Libert}, A.~S., \& {Lega}, E. 2013, \aap, 555,
  A124, \dodoi{10.1051/0004-6361/201220310}

\bibitem[{{Calvet} {et~al.}(2002){Calvet}, {D'Alessio}, {Hartmann}, {Wilner},
  {Walsh}, \& {Sitko}}]{Calvet2002}
{Calvet}, N., {D'Alessio}, P., {Hartmann}, L., {et~al.} 2002, \apj, 568, 1008,
  \dodoi{10.1086/339061}

\bibitem[{{Calvet} {et~al.}(2005){Calvet}, {D'Alessio}, {Watson},
  {Franco-Hern{\'a}ndez}, {Furlan}, {Green}, {Sutter}, {Forrest}, {Hartmann},
  {Uchida}, {Keller}, {Sargent}, {Najita}, {Herter}, {Barry}, \&
  {Hall}}]{Calvet2005}
{Calvet}, N., {D'Alessio}, P., {Watson}, D.~M., {et~al.} 2005, \apjl, 630,
  L185, \dodoi{10.1086/491652}

\bibitem[{{Dawson} \& {Johnson}(2018)}]{Dawson2018}
{Dawson}, R.~I., \& {Johnson}, J.~A. 2018, \araa, 56, 175,
  \dodoi{10.1146/annurev-astro-081817-051853}

\bibitem[{{Dodson-Robinson} \& {Salyk}(2011)}]{DodsonRobinson2011}
{Dodson-Robinson}, S.~E., \& {Salyk}, C. 2011, \apj, 738, 131,
  \dodoi{10.1088/0004-637X/738/2/131}

\bibitem[{{Donati} {et~al.}(2020){Donati}, {Bouvier}, {Alencar}, {Moutou},
  {Malo}, {Takami}, {M{\'e}nard}, {Dougados}, {Hussain}, \& {Matysse
  Collaboration}}]{Donati2020}
{Donati}, J.~F., {Bouvier}, J., {Alencar}, S.~H., {et~al.} 2020, \mnras, 491,
  5660, \dodoi{10.1093/mnras/stz3368}

\bibitem[{{Dong}(2015)}]{Dong2015}
{Dong}, R. 2015, \apj, 810, 6, \dodoi{10.1088/0004-637X/810/1/6}

\bibitem[{{Dong} \& {Fung}(2017)}]{Dong2017}
{Dong}, R., \& {Fung}, J. 2017, \apj, 835, 146,
  \dodoi{10.3847/1538-4357/835/2/146}

\bibitem[{{Dong} {et~al.}(2012{\natexlab{a}}){Dong}, {Rafikov}, {Zhu},
  {Hartmann}, {Whitney}, {Brandt}, {Muto}, {Hashimoto}, {Grady}, {Follette},
  {Kuzuhara}, {Tanii}, {Itoh}, {Thalmann}, {Wisniewski}, {Mayama}, {Janson},
  {Abe}, {Brandner}, {Carson}, {Egner}, {Feldt}, {Goto}, {Guyon}, {Hayano},
  {Hayashi}, {Hayashi}, {Henning}, {Hodapp}, {Honda}, {Inutsuka}, {Ishii},
  {Iye}, {Kandori}, {Knapp}, {Kudo}, {Kusakabe}, {Matsuo}, {McElwain},
  {Miyama}, {Morino}, {Moro-Martin}, {Nishimura}, {Pyo}, {Suto}, {Suzuki},
  {Takami}, {Takato}, {Terada}, {Tomono}, {Turner}, {Watanabe}, {Yamada},
  {Takami}, {Usuda}, \& {Tamura}}]{dong12cavity}
{Dong}, R., {Rafikov}, R., {Zhu}, Z., {et~al.} 2012{\natexlab{a}}, \apj, 750,
  161, \dodoi{10.1088/0004-637X/750/2/161}

\bibitem[{{Dong} {et~al.}(2012{\natexlab{b}}){Dong}, {Hashimoto}, {Rafikov},
  {Zhu}, {Whitney}, {Kudo}, {Muto}, {Brandt}, {McClure}, {Wisniewski}, {Abe},
  {Brandner}, {Carson}, {Egner}, {Feldt}, {Goto}, {Grady}, {Guyon}, {Hayano},
  {Hayashi}, {Hayashi}, {Henning}, {Hodapp}, {Ishii}, {Iye}, {Janson},
  {Kandori}, {Knapp}, {Kusakabe}, {Kuzuhara}, {Kwon}, {Matsuo}, {McElwain},
  {Miyama}, {Morino}, {Moro-Martin}, {Nishimura}, {Pyo}, {Serabyn}, {Suto},
  {Suzuki}, {Takami}, {Takato}, {Terada}, {Thalmann}, {Tomono}, {Turner},
  {Watanabe}, {Yamada}, {Takami}, {Usuda}, \& {Tamura}}]{Dong2012}
{Dong}, R., {Hashimoto}, J., {Rafikov}, R., {et~al.} 2012{\natexlab{b}}, \apj,
  760, 111, \dodoi{10.1088/0004-637X/760/2/111}

\bibitem[{{Dunhill} {et~al.}(2013){Dunhill}, {Alexander}, \&
  {Armitage}}]{Dunhill2013}
{Dunhill}, A.~C., {Alexander}, R.~D., \& {Armitage}, P.~J. 2013, \mnras, 428,
  3072, \dodoi{10.1093/mnras/sts254}

\bibitem[{{Espaillat} {et~al.}(2011){Espaillat}, {Furlan}, {D'Alessio},
  {Sargent}, {Nagel}, {Calvet}, {Watson}, \& {Muzerolle}}]{Espaillat2011}
{Espaillat}, C., {Furlan}, E., {D'Alessio}, P., {et~al.} 2011, \apj, 728, 49,
  \dodoi{10.1088/0004-637X/728/1/49}

\bibitem[{{Espaillat} {et~al.}(2014){Espaillat}, {Muzerolle}, {Najita},
  {Andrews}, {Zhu}, {Calvet}, {Kraus}, {Hashimoto}, {Kraus}, \&
  {D'Alessio}}]{Espaillat2014}
{Espaillat}, C., {Muzerolle}, J., {Najita}, J., {et~al.} 2014, in Protostars
  and Planets VI, ed. H.~{Beuther}, R.~S. {Klessen}, C.~P. {Dullemond}, \&
  T.~{Henning}, 497, \dodoi{10.2458/azu\_uapress\_9780816531240-ch022}

\bibitem[{{Flagg} {et~al.}(2019){Flagg}, {Johns-Krull}, {Nofi}, {Llama},
  {Prato}, {Sullivan}, {Jaffe}, \& {Mace}}]{Flagg2019}
{Flagg}, L., {Johns-Krull}, C.~M., {Nofi}, L., {et~al.} 2019, \apjl, 878, L37,
  \dodoi{10.3847/2041-8213/ab276d}

\bibitem[{{Flaherty} {et~al.}(2012){Flaherty}, {Muzerolle}, {Rieke},
  {Gutermuth}, {Balog}, {Herbst}, {Megeath}, \& {Kun}}]{Flaherty2012}
{Flaherty}, K.~M., {Muzerolle}, J., {Rieke}, G., {et~al.} 2012, \apj, 748, 71,
  \dodoi{10.1088/0004-637X/748/1/71}

\bibitem[{{Fung}(2015)}]{Fung2015}
{Fung}, J. 2015, PhD thesis, University of Toronto, Canada

\bibitem[{{Fung} \& {Chiang}(2016)}]{Fung2016}
{Fung}, J., \& {Chiang}, E. 2016, \apj, 832, 105,
  \dodoi{10.3847/0004-637X/832/2/105}

\bibitem[{{Fung} {et~al.}(2014){Fung}, {Shi}, \& {Chiang}}]{Fung2014}
{Fung}, J., {Shi}, J.-M., \& {Chiang}, E. 2014, \apj, 782, 88,
  \dodoi{10.1088/0004-637X/782/2/88}

\bibitem[{{Guilloteau} {et~al.}(2014){Guilloteau}, {Simon}, {Pi{\'e}tu}, {Di
  Folco}, {Dutrey}, {Prato}, \& {Chapillon}}]{Guilloteau2014}
{Guilloteau}, S., {Simon}, M., {Pi{\'e}tu}, V., {et~al.} 2014, \aap, 567, A117,
  \dodoi{10.1051/0004-6361/201423765}

\bibitem[{{Isella} \& {Natta}(2005)}]{Isella2005}
{Isella}, A., \& {Natta}, A. 2005, \aap, 438, 899,
  \dodoi{10.1051/0004-6361:20052773}

\bibitem[{{Johns-Krull} {et~al.}(2016){Johns-Krull}, {McLane}, {Prato},
  {Crockett}, {Jaffe}, {Hartigan}, {Beichman}, {Mahmud}, {Chen}, {Skiff},
  {Cauley}, {Jones}, \& {Mace}}]{JohnsKrull2016}
{Johns-Krull}, C.~M., {McLane}, J.~N., {Prato}, L., {et~al.} 2016, \apj, 826,
  206, \dodoi{10.3847/0004-637X/826/2/206}

\bibitem[{{Keppler} {et~al.}(2018){Keppler}, {Benisty}, {M{\"u}ller},
  {Henning}, {van Boekel}, {Cantalloube}, {Ginski}, {van Holstein}, {Maire},
  {Pohl}, {Samland}, {Avenhaus}, {Baudino}, {Boccaletti}, {de Boer},
  {Bonnefoy}, {Chauvin}, {Desidera}, {Langlois}, {Lazzoni}, {Marleau},
  {Mordasini}, {Pawellek}, {Stolker}, {Vigan}, {Zurlo}, {Birnstiel},
  {Brandner}, {Feldt}, {Flock}, {Girard}, {Gratton}, {Hagelberg}, {Isella},
  {Janson}, {Juhasz}, {Kemmer}, {Kral}, {Lagrange}, {Launhardt}, {Matter},
  {M{\'e}nard}, {Milli}, {Molli{\`e}re}, {Olofsson}, {P{\'e}rez}, {Pinilla},
  {Pinte}, {Quanz}, {Schmidt}, {Udry}, {Wahhaj}, {Williams}, {Buenzli},
  {Cudel}, {Dominik}, {Galicher}, {Kasper}, {Lannier}, {Mesa}, {Mouillet},
  {Peretti}, {Perrot}, {Salter}, {Sissa}, {Wildi}, {Abe}, {Antichi},
  {Augereau}, {Baruffolo}, {Baudoz}, {Bazzon}, {Beuzit}, {Blanchard}, {Brems},
  {Buey}, {De Caprio}, {Carbillet}, {Carle}, {Cascone}, {Cheetham}, {Claudi},
  {Costille}, {Delboulb{\'e}}, {Dohlen}, {Fantinel}, {Feautrier}, {Fusco},
  {Giro}, {Gluck}, {Gry}, {Hubin}, {Hugot}, {Jaquet}, {Le Mignant}, {Llored},
  {Madec}, {Magnard}, {Martinez}, {Maurel}, {Meyer}, {M{\"o}ller-Nilsson},
  {Moulin}, {Mugnier}, {Orign{\'e}}, {Pavlov}, {Perret}, {Petit}, {Pragt},
  {Puget}, {Rabou}, {Ramos}, {Rigal}, {Rochat}, {Roelfsema}, {Rousset}, {Roux},
  {Salasnich}, {Sauvage}, {Sevin}, {Soenke}, {Stadler}, {Suarez}, {Turatto}, \&
  {Weber}}]{Keppler2018}
{Keppler}, M., {Benisty}, M., {M{\"u}ller}, A., {et~al.} 2018, \aap, 617, A44,
  \dodoi{10.1051/0004-6361/201832957}

\bibitem[{{Kim} {et~al.}(1994){Kim}, {Martin}, \& {Hendry}}]{Kim1994}
{Kim}, S.-H., {Martin}, P.~G., \& {Hendry}, P.~D. 1994, \apj, 422, 164,
  \dodoi{10.1086/173714}

\bibitem[{{Kley} \& {Dirksen}(2006)}]{Kley2006}
{Kley}, W., \& {Dirksen}, G. 2006, \aap, 447, 369,
  \dodoi{10.1051/0004-6361:20053914}

\bibitem[{{Kraus} {et~al.}(2013){Kraus}, {Ireland}, {Sitko}, {Monnier},
  {Calvet}, {Espaillat}, {Grady}, {Harries}, {H{\"o}nig}, {Russell},
  {Swearingen}, {Werren}, \& {Wilner}}]{Kraus2013}
{Kraus}, S., {Ireland}, M.~J., {Sitko}, M.~L., {et~al.} 2013, \apj, 768, 80,
  \dodoi{10.1088/0004-637X/768/1/80}

\bibitem[{{Manger} {et~al.}(2021){Manger}, {Pfeil}, \& {Klahr}}]{Manger2021}
{Manger}, N., {Pfeil}, T., \& {Klahr}, H. 2021, \mnras,
  \dodoi{10.1093/mnras/stab2599}

\bibitem[{{McClure} {et~al.}(2013{\natexlab{a}}){McClure}, {D'Alessio},
  {Calvet}, {Espaillat}, {Hartmann}, {Sargent}, {Watson}, {Ingleby}, \&
  {Hern{\'a}ndez}}]{McClure2013}
{McClure}, M.~K., {D'Alessio}, P., {Calvet}, N., {et~al.} 2013{\natexlab{a}},
  \apj, 775, 114, \dodoi{10.1088/0004-637X/775/2/114}

\bibitem[{{McClure} {et~al.}(2013{\natexlab{b}}){McClure}, {Calvet},
  {Espaillat}, {Hartmann}, {Hern{\'a}ndez}, {Ingleby}, {Luhman}, {D'Alessio},
  \& {Sargent}}]{McClure2013b}
{McClure}, M.~K., {Calvet}, N., {Espaillat}, C., {et~al.} 2013{\natexlab{b}},
  \apj, 769, 73, \dodoi{10.1088/0004-637X/769/1/73}

\bibitem[{{Miranda} \& {Rafikov}(2020)}]{MirandaRafikov2020}
{Miranda}, R., \& {Rafikov}, R.~R. 2020, \apj, 892, 65,
  \dodoi{10.3847/1538-4357/ab791a}

\bibitem[{{Muley} {et~al.}(2021){Muley}, {Dong}, \& {Fung}}]{Muley2021}
{Muley}, D., {Dong}, R., \& {Fung}, J. 2021, arXiv e-prints, arXiv:2107.06323.
\newblock \doarXiv{2107.06323}

\bibitem[{{Muley} {et~al.}(2019){Muley}, {Fung}, \& {van der
  Marel}}]{Muley2019}
{Muley}, D., {Fung}, J., \& {van der Marel}, N. 2019, \apjl, 879, L2,
  \dodoi{10.3847/2041-8213/ab24d0}

\bibitem[{{M{\"u}ller} {et~al.}(2018){M{\"u}ller}, {Keppler}, {Henning},
  {Samland}, {Chauvin}, {Beust}, {Maire}, {Molaverdikhani}, {van Boekel},
  {Benisty}, {Boccaletti}, {Bonnefoy}, {Cantalloube}, {Charnay}, {Baudino},
  {Gennaro}, {Long}, {Cheetham}, {Desidera}, {Feldt}, {Fusco}, {Girard},
  {Gratton}, {Hagelberg}, {Janson}, {Lagrange}, {Langlois}, {Lazzoni}, {Ligi},
  {M{\'e}nard}, {Mesa}, {Meyer}, {Molli{\`e}re}, {Mordasini}, {Moulin},
  {Pavlov}, {Pawellek}, {Quanz}, {Ramos}, {Rouan}, {Sissa}, {Stadler}, {Vigan},
  {Wahhaj}, {Weber}, \& {Zurlo}}]{Muller2018}
{M{\"u}ller}, A., {Keppler}, M., {Henning}, T., {et~al.} 2018, \aap, 617, L2,
  \dodoi{10.1051/0004-6361/201833584}

\bibitem[{{Papaloizou} {et~al.}(2001){Papaloizou}, {Nelson}, \&
  {Masset}}]{Papaloizou2001}
{Papaloizou}, J.~C.~B., {Nelson}, R.~P., \& {Masset}, F. 2001, \aap, 366, 263,
  \dodoi{10.1051/0004-6361:20000011}

\bibitem[{{Rafikov}(2016)}]{Rafikov2016}
{Rafikov}, R.~R. 2016, \apj, 831, 122, \dodoi{10.3847/0004-637X/831/2/122}

\bibitem[{{Ragusa} {et~al.}(2020){Ragusa}, {Alexander}, {Calcino}, {Hirsh}, \&
  {Price}}]{Ragusa2020}
{Ragusa}, E., {Alexander}, R., {Calcino}, J., {Hirsh}, K., \& {Price}, D.~J.
  2020, \mnras, 499, 3362, \dodoi{10.1093/mnras/staa2954}

\bibitem[{{Ragusa} {et~al.}(2018){Ragusa}, {Rosotti}, {Teyssandier}, {Booth},
  {Clarke}, \& {Lodato}}]{Ragusa2018}
{Ragusa}, E., {Rosotti}, G., {Teyssandier}, J., {et~al.} 2018, \mnras, 474,
  4460, \dodoi{10.1093/mnras/stx3094}

\bibitem[{{Sallum} {et~al.}(2015){Sallum}, {Follette}, {Eisner}, {Close},
  {Hinz}, {Kratter}, {Males}, {Skemer}, {Macintosh}, {Tuthill}, {Bailey},
  {Defr{\`e}re}, {Morzinski}, {Rodigas}, {Spalding}, {Vaz}, \&
  {Weinberger}}]{Sallum2015}
{Sallum}, S., {Follette}, K.~B., {Eisner}, J.~A., {et~al.} 2015, \nat, 527,
  342, \dodoi{10.1038/nature15761}

\bibitem[{{Teyssandier} \& {Lai}(2020)}]{Teyssandier2020}
{Teyssandier}, J., \& {Lai}, D. 2020, \mnras, 495, 3920,
  \dodoi{10.1093/mnras/staa1363}

\bibitem[{{Teyssandier} \& {Ogilvie}(2017)}]{Teyssandier2017}
{Teyssandier}, J., \& {Ogilvie}, G.~I. 2017, \mnras, 467, 4577,
  \dodoi{10.1093/mnras/stx426}

\bibitem[{{van der Marel} \& {Mulders}(2021)}]{vanDerMarel2021b}
{van der Marel}, N., \& {Mulders}, G.~D. 2021, \aj, 162, 28,
  \dodoi{10.3847/1538-3881/ac0255}

\bibitem[{{van der Marel} {et~al.}(2018){van der Marel}, {Williams}, {Ansdell},
  {Manara}, {Miotello}, {Tazzari}, {Testi}, {Hogerheijde}, {Bruderer}, {van
  Terwisga}, \& {van Dishoeck}}]{VanDerMarel2018}
{van der Marel}, N., {Williams}, J.~P., {Ansdell}, M., {et~al.} 2018, \apj,
  854, 177, \dodoi{10.3847/1538-4357/aaaa6b}

\bibitem[{{Whitney} {et~al.}(2013){Whitney}, {Robitaille}, {Bjorkman}, {Dong},
  {Wolff}, {Wood}, \& {Honor}}]{Whitney2013}
{Whitney}, B.~A., {Robitaille}, T.~P., {Bjorkman}, J.~E., {et~al.} 2013, \apjs,
  207, 30, \dodoi{10.1088/0067-0049/207/2/30}

\bibitem[{{Zhang} {et~al.}(2014){Zhang}, {Isella}, {Carpenter}, \&
  {Blake}}]{Zhang2014}
{Zhang}, K., {Isella}, A., {Carpenter}, J.~M., \& {Blake}, G.~A. 2014, \apj,
  791, 42, \dodoi{10.1088/0004-637X/791/1/42}

\bibitem[{{Zhu} {et~al.}(2012){Zhu}, {Nelson}, {Dong}, {Espaillat}, \&
  {Hartmann}}]{Zhu2012}
{Zhu}, Z., {Nelson}, R.~P., {Dong}, R., {Espaillat}, C., \& {Hartmann}, L.
  2012, \apj, 755, 6, \dodoi{10.1088/0004-637X/755/1/6}

\bibitem[{{Zhu} {et~al.}(2011){Zhu}, {Nelson}, {Hartmann}, {Espaillat}, \&
  {Calvet}}]{Zhu2011}
{Zhu}, Z., {Nelson}, R.~P., {Hartmann}, L., {Espaillat}, C., \& {Calvet}, N.
  2011, \apj, 729, 47, \dodoi{10.1088/0004-637X/729/1/47}

\end{thebibliography}



\end{document}